\documentclass[11pt]{article}
\usepackage{amsthm,amsmath}
\usepackage{graphicx,psfrag,epsf}
\usepackage{xcolor}
\usepackage{subcaption}
\usepackage{enumerate}
\usepackage{natbib}
\usepackage[title]{appendix}
\usepackage{setspace}
\usepackage{enumitem}

\usepackage{url} 

\newcommand{\blind}{0}

\numberwithin{equation}{section}
\theoremstyle{plain}

\begin{document}

\def\spacingset#1{\renewcommand{\baselinestretch}%
{#1}\small\normalsize} \spacingset{1}

\def\B0{\mbox{\boldmath $0$}}
\def\Ba{\mbox{\boldmath $A$}}
\def\Bb{\mbox{\boldmath $B$}}
\def\Bc{\mbox{\boldmath $C$}}
\def\Bd{\mbox{\boldmath $D$}}
\def\Bl{\mbox{\boldmath $L$}}
\def\Bn{\mbox{\boldmath $N$}}
\def\Br{\mbox{\boldmath $R$}}
\def\Bt{\mbox{\boldmath $T$}}
\def\Bv{\mbox{\boldmath $V$}}
\def\Bx{\mbox{\boldmath $X$}}
\def\By{\mbox{\boldmath $Y$}}
\def\Bz{\mbox{\boldmath $Z$}}

\def\btheta{\mbox{\boldmath $\theta$}}
\def\htheta{\mbox{$\hat{\theta}$}}

\def\balpha{\mbox{\boldmath $\alpha$}}
\def\bbeta{\mbox{\boldmath $\beta$}}
\def\bdelta{\mbox{\boldmath $\delta$}}
\def\bepsilon{\mbox{\boldmath $\epsilon$}}
\def\bGamma{\mbox{\boldmath $\Gamma$}}
\def\blambda{\mbox{\boldmath $\lambda$}}
\def\bLambda{\mbox{\boldmath $\Lambda$}}
\def\bmu{\mbox{\boldmath $\mu$}}
\def\bSigma{\mbox{\boldmath $\Sigma$}}
\def\btau{\mbox{\boldmath $\tau$}}
\def\btheta{\mbox{\boldmath $\theta$}}


\if0\blind
{
  \title{
  	On Assessing Overall Survival (OS) in Oncology Studies
  	}
  \author{
  	Jason C. Hsu, Department of Statistics, The Ohio State University\\
  }
  \maketitle
} \fi

\if1\blind
{
  \bigskip
  \bigskip
  \bigskip
  \begin{center}
    {\LARGE\bf Title}
\end{center}
  \medskip
} \fi


\noindent%
{\it Keywords:}  Hazard Ratio; Logic-respecting; log-Rank test; Type I error rate
\vfill

\spacingset{1.45} 

\section{Efficacy measures should be Logic-respecting}

Targeted therapies are common in oncology.  
A targeted therapy may have differential effect in different biomarker-driven subgroups.  
For simplicity, consider the situation that there is a marker-positive subgroup and its complementary marker-negative subgroup.  
Medically, an efficacy measure should be \emph{Logic-respecting}, with efficacy in the overall population being between efficacy in the marker-positive $g^+$ subgroup and the marker-negative $g^-$ subgroup.  

\begin{figure}[h!]
	\centering
	\includegraphics[width=0.7\linewidth]{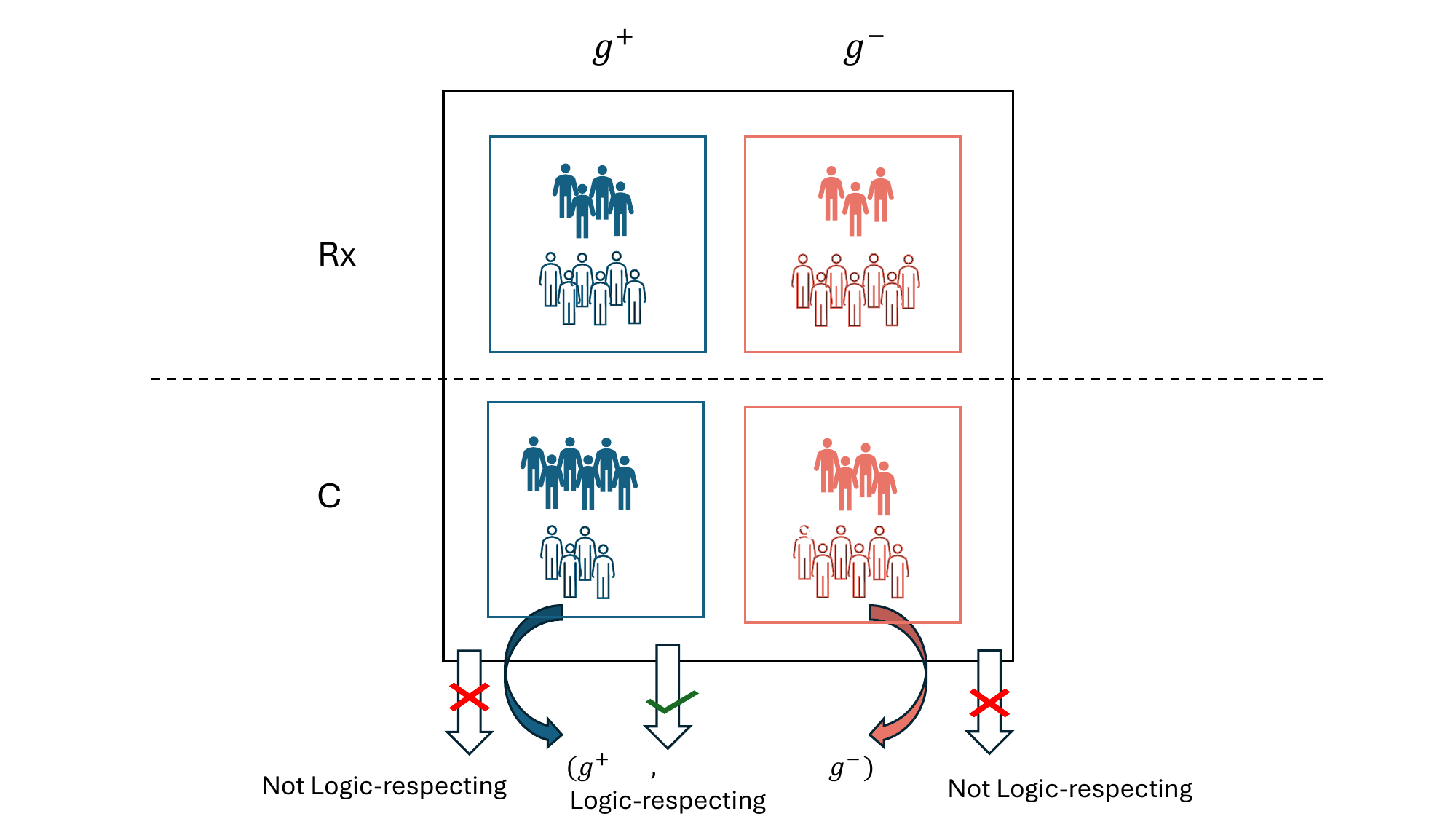}
	\caption[Logic-respecting efficacy measure]{Using solid figures to represent \textit{responders} and outlined figures to represent \textit{non-responders}, this plot illustrates Relative Response (RR) is Logic-respecting, with efficacy in $g^+$ being 2/3, efficacy in $g^+$ being 3/4, and efficacy in the overall population being 7/10 which is between 2/3 and 3/4 (closer to 2/3 than to 3/4).  However, we will show that Hazard Ratio (HR) is not Logic-respecting and that may lead to incorrect patient-targeting.}
	\label{fig:logicRespecting}
\end{figure}


\subsection{Logic-respecting is more relevant than being Collapsible}

The Logic-respecting concept was first introduced in \cite{Ding&Lin&Hsu(2016)}, who proved that Time Ratio (TR) such as ratio of median survival times is Logic-respecting.  
As explained in the highly influential Causal Inference paper \cite{ColnetEtAl(2023)}, this concept of \textit{Logicality} is distinct from Collapsibility, one version of which is efficacy in the overall population can be expressed as some weighted function of efficacy in the subgroups.  

As an example of a Logic-respecting efficacy measure which is not Collapsible, TR is Logic-respecting but TR in the overall population cannot be explicitly expressed as a weighted function of efficacy in the subgroups, because median survival time (say) is the \emph{implicit} solution to an equation that the survival probability \emph{at that time} equals $50\%$.  

As an example of an efficacy measure which is Collapsible but not Logic-respecting, Living Longer Probability (LLP) as defined in \cite{LiuWangTianHsu(2022)} is Collapsible because LLP in the overall population is a linear combination of four LLPs: LLP of marker-\emph{positive} patients treated with $Rx$ versus marker-\emph{positive} patients treated with $C$, LLP of marker-\emph{negative} patients treated with $Rx$ versus marker-\emph{negative} patients treated with $C$, LLP of marker-\emph{positive} patients treated with $Rx$ versus marker-\emph{negative} patients treated with $C$, and LLP of marker-\emph{negative} patients treated with $Rx$ versus marker-\emph{positive} patients treated with $C$.  
However, LLP is not Logic-respecting bcause as shown in \cite{LiuWangTianHsu(2022)} it is equivalent to Hazard Ratio (HR) and HR is not Logic-respecting as we explain below.  

\section{Hazard Ratio (HR) is not Logic-respecting}\label{sec.HRnotLogical}

To understand Hazard Ratio (HR), it is imperative to know Cox’s original motivation for inventing it, which is on Page 23 of \cite{Cox&Oaks(1984)}.  
The idea is simple: to have a single parameter ordering a family of survival curves, they cannot differ by location shifts with some starting at time zero while others starting at other times.  
They cannot differ by scaling as all survival functions must start with 100\% probability and end with zero probability.  
Survival probability of a reference distribution are numbers between zero and one at all time points.  
Squaring the distribution, for example, makes all the probabilities lower, and Cox defines that as HR = 2.  
Taking square root of the reference distribution makes all the probabilities higher, and Cox defines that as HR = 1/2, for example.  
This family of survival distributions in which each is the reference distribution raised to an exponent is called a Lehmann family, with Proportional Hazards (PH), and the exponent to which the reference distribution is raised is what Cox defines as the HR relative to the reference survival distribution.  
From this original definition, it is easy to derive as \cite{LiuWangTianHsu(2022)} did that HR is the \emph{\textbf{Odds} of Living Shorter}.  
For example, $Rx\!:\!C$ having an HR of 2/3 means the odds that a randomly picked patient given $Rx$ would have a shorter survival time then another randomly picked patient given $C$ is 2-to-3 (i.e., the \emph{Odds of Living Longer} is 3-to-2).  

\subsection{Hazard Ratio as efficacy measure can target wrong patients}

Gandara et al. (2018) assessed blood-based Tumor Mutational Burden (bTMB) as a biomarker to select patients to receive atezolizumab instead of docetaxel in Non-Small-Cell Lung Cancer (NSCLC).  
Using hazard ratio (HR) as efficacy measure for OS in the OAK study, 
they reported patients with \emph{higher} bTMB values benefit more from atezolizumab relative to docetaxel.  
Analyzing the publicly accessible patient-level data, \cite{LiuWangKilHsu(2022)} were surprised to find patients with \emph{lower} bTMB values also appear to benefit more from atezolizumab relative to docetaxel.  

\begin{figure}[h!]
	\centering
	\includegraphics[width=0.7\linewidth]{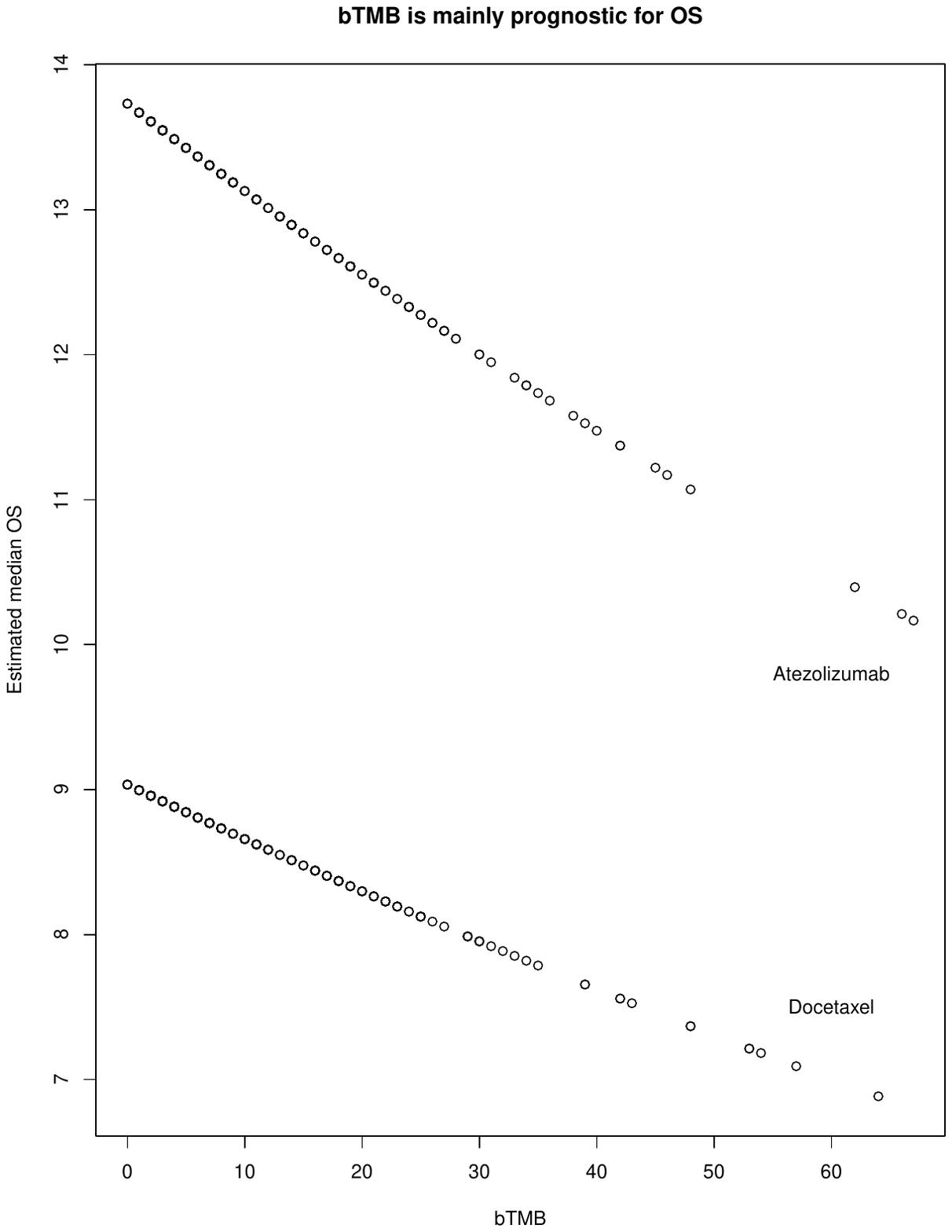}
	\caption{With the estimated median overall survival (OS) times being nearly parallel between atezolizumab and docetaxel across the range of bTMB values, it appears that bTMB is purely prognostic for OS as defined in Section 5 of \cite{LiuWangKilHsu(2022)}.}
	\label{fig:Oak_OS_prognostic}
\end{figure}

After careful modeling of the OAK study data, Figure \ref{fig:Oak_OS_prognostic} (which is Figure 7 in \cite{LiuWangKilHsu(2022)}) shows, patients benefit \emph{equally} from atezolizumab relative to docetaxel in terms of predicted survival time, regardless of bTMB value.  

Why using Hazard Ratio (HR) as efficacy measure created the illusion that patients with \emph{higher} or \emph{lower} bTMB benefit more from atezolizumab over docetaxel is because, as shown in Theorem 2 of \cite{LiuWangKilHsu(2022)}, 
HR in the overall population (``average'' HR) is a \emph{dilution} 
of HR in its subgroups.  
Using HR as efficacy measure can thus deprive patients from receiving beneficial therapies.  

\subsection{Computer stratified analysis results have been misleading}

For ratio efficacy (Relative Risk, Odds Ratio, Hazard Ratio), since the year 2008, computer packages\footnote{Take the analysis of Hazard Ratio (HR) as an example, JMP Analyze's \texttt{Fit Proportional Hazard} platform, R \texttt{multcomp} package's \texttt{glht} function, and SAS' \texttt{Proc PHreg} all have the same issue.} have been incorrectly computing stratified efficacy for the overall population.  

For example, as shown in Table \ref{tab:Watermelon}, for the (same) OAK data set, computer packages senselessly give widely different stratified HR for the overall population, just depending on how the population is stratified in the analysis.  
This is because computer packages calculate stratified HR as an average of the logarithms of the conditional HR like in equation (\ref{eq.MixingSex}), ignoring the Prognostic effect which was precisely defined in Section 5 of \cite{LiuWangKilHsu(2022)}.  
For example, with Sex as a stratification factor, HR is 0.521 for Female and 0.983 for Male, ignoring the Prognostic effect gives (for a perfectly balanced population) an average HR of 
\begin{equation}
	e^{0.5 \times (log(0.521) + 0.5 \times log(0.983))} = 0.716. \label{eq.MixingSex}
\end{equation}
Since the prognostic effect of Decetaxel differ for Sex, Histology, KRAS, and EGFR, so do the computer package results, even though the results should all remain around 0.767.  
\begin{table}[htb!]
	\centering
	\begin{tabular}{|c|c|c|c|c|}
		\hline
		& Sex & Histology & KRAS & EGFR \\
		\hline
		 JMP/R/SAS stratified analysis results & \textcolor{red}{0.716} & 0.770 & \textcolor{red}{0.738} & \textcolor{red}{0.972} \\
		\hline
		Subgroup Mixable Estimation (SME) results & 0.768 & 0.758 & 0.774 & 0.771 \\
		\hline
	\end{tabular}
	\caption{Illustration of misleading stratified HR analysis results from computer packages using the subset of the OAK data with no missing values for any of the stratification factors Sex, Histology (squamous vs. non-squamous), KRAS, and EGFR, with the marginal HR being approximately 0.767.  For ratio efficacy (Relative Risk, Odds Ratio, Hazard Ratio), computer packages mix logarithms of conditional efficacy, ignoring the \emph{Prognostic} effect.  As the prognostic effect of the Control treatment (Docetaxel) varies for Female:Male, Squamous:Non-squamous, KRAS-mutated:Wild-type, EGFR-mutated:Wild-type, stratified HR given by computer packages senselessly varies as well, as indicted in \textcolor{red}{red}.  However, Subgroup Mixable Estimation (SME) following equation (2) in \cite{LiuWangTianHsu(2022)} consistently produces correct (albeit not necessarily Logic-respecting) stratified HR results.}\label{tab:Watermelon}
\end{table}

Sections 3.2 and A.3 of \cite{LiuEtAl(2023)} shows that, with or without this computer package error, using HR as efficacy measure can lead to wrong targeting of patients in choosing a cut-point of a biomarker (such as bTMB) to target patients with.  

This oversight was due to the seemingly plausible but deadly incorrect thinking that, 
\begin{itemize}[noitemsep, topsep=0pt]
	\item similar to difference efficacy, ratio efficacy in the overall population is a linear combination of (some form of) efficacy in the subgroups;  
	\item ``\emph{log} of a \textit{ratio} is a \textit{difference} of \emph{logs}''.
\end{itemize}
A lesson from the SME principle is this too naive thinking overlooks that the order of operations of \emph{addition} first and then \emph{division} cannot be reversed.  
\cite{LiuWangTianHsu(2022)} shows the way of using SME to correct this patient-harming error in the computer packages, regardless of whether the efficacy measure is Logic-respecting or not.  


\section{Log-Rank test has inflated incorrect decision rate}

A non-sensible practice in the statistical analysis of survival data in oncology studies is to report confidence intervals for Hazard Ratio (HR) from the Wald test in a Cox Proportional Hazard (PH) model, but report p-values from the (Score) log-Rank test.  

The null hypothesis being tested by the (Score) log-Rank test is the survival functions under $ Rx $ and $ C $ are exactly equal at {\em all} time points, 
which can also be thought of as $Rx$ and $C$ having the same (population) quantile survival times (e.g., median survival times) for all quantiles.  
That this Null-null hypothesis\footnote{A Null-null hypothesis is one such that it is surely to be false.} is false is a given, with no statistical testing required to established it.\footnote{As \cite{Tukey(1992)} said, ``provided we measure to enough decimal places, no two `treatments' ever have identically the same long-run value''.}  
As sample size gets bigger, the p-value of the log-Rank test of a Null-null will get smaller, leading surely to an eventual rejection.  
Logically, upon rejection of the (Score) log-Rank test, one can only infer ``the survival functions given $ Rx $ and $ C $ differ at some time point'', which is not an actionable inference.  

So suppose the decision-making process is, \emph{if} the log-Rank test rejects, \emph{then} one looks at the  Kaplan–Meier (KM) curves and declares $Rx$ and $C$ to be different wherever one seemingly sees clinically meaningful separation.  
This is classic Texas Sharpshooting: 
\begin{enumerate}[noitemsep, topsep=0pt]
	\item Randomly fire shots at a barn.
	\item Identified a cluster of bullet holes that appear close together.  
	\item Draw a target around the cluster, claiming sharp-shooting skill.  
\end{enumerate}

\noindent Such claims are unlikely to be true, because all the other bullet holes are ignored.  

So, instead of \textit{post hoc} inference, suppose there is a predefined decision-making plan.  
For outcome measures that are not time-to-event, it has long been recognized that Type I error rate control testing the null hypothesis that \emph{all} its component nulls are true, called \textit{weak} Error Rate Familywise (ERFw) control, is inadequate, because it may not translate into control of Incorrect 
Decision rate.  
For example, with multiple co-primary endpoints, \cite{Hsu&Berger(1999)} showed the scenario that has the highest probability for the standard pairwise method to incorrectly infer that there is efficacy in both endpoints is \emph{not} when there is no efficacy in the primary endpoint \emph{and} there is no efficacy in the secondary endpoint.  
We now show that Type I error rate control by a 2-sided (score) log-Rank test offers no protection against the rate of \emph{making incorrect decisions}.  

Consider making decision by first conducting a level-$\alpha$ (score) log-Rank test and, upon rejection, whichever treatment arm has the longer estimated median survival time, infer that treatment has longer median survival time than the other treatment for the overall population.  
Consider a scenario where \textit{one} but \emph{not all} of the equality nulls of expected survival times are true, specifically that the median survival times are equal between $ Rx $ and $ C $ in the overall population, but other survival time quantiles may differ, specifically, there are subgroups so that in the $ g^+ $ subgroup patients given $Rx$ do better than those given $C$ but the reverse is true in the $ g^- $ patients. 
Of course, asserting either $Rx$ or $C$ has longer \emph{median} survival time constitutes a \emph{directional error}.   
\cite{Liu&Tian&Hsu(2021)} conducted a simulation study to see what is the probability that this decision-making process would make an \textit{incorrect directional} medical claim.  

For their simulation, the median survival time for overall population was set to be 8 months under both $Rx$ and $C$.  
Data is generated from Weibull distributions, with shape parameter values of 1.05 and 1.20 for the $g^-$ and $g^+$ subgroups respectively.  
They generated data sets with sample size 1000, equally randomized to $Rx$ and $C$, with a prevalence of $50\%$ for each of the $g^-$ and $g^+$ subgroups, without censoring.  
For $g^+$ patients, the median survival times are 12 months and 6 months given $Rx$ and $C$ respectively.
In their simulation, they made use of the fact that, within each treatment arm and at each time point, the survival probability in the overall population is a mixture of survival probabilities in the $g^-$ and $g^+$ subgroups, and median survival times in the overall population and in the $g^+$ subgroup determine the scale parameter values in the $g^-$ subgroup.  
Setting the level of the 2-sided log-Rank test at $5\%$, they found the percent of times it rejects was 304 times out of the 1000 Weibull data sets simulated.  

With the true median survival times under $Rx$ and $C$ being the same, inferring either $Rx$ or $C$ as having longer median survival time is a directional error, an incorrect decision.  
For a $5\%$ 2-sided test which rejects when an equal-tailed $95\%$ \emph{confidence interval} does not cover the null, this incorrect decision rate would be no more than $\mathbf{2.5}\%$.  
On the contrary, for the (Score) log-Rank test, since the sum of the two possible directional error rates is estimated to exceed $30\%$, at least one of the two directional error rates exceeds $\mathbf{15}\%$.  

The (Score) log-Rank test is popular because it is perceived to be more ``powerful'' than the Wald test, a \emph{mis}perception due to including the probability of \textit{rejecting for wrong reasons} in ``power".  
Thus, for time-to-event outcomes, we urge a fundamental re-assessment of the concept of (regulatory) Type I error rate control, vis-$ \grave{a} $-vis the (Score) log-Rank test.  

\section{Time Ratio (TR) can be recommended}

Overall Survival (OS) itself is measured in terms of \emph{time}, so it seems natural to assess treatment efficacy on OS in terms of \textit{time}.  

To measure Clinical Benefit of cancer treatments, the value framework of the American Society of Clinical Oncology (ASCO), \cite{ASCO(2015)}, lists \textit{fractional improvement in median OS} as the first item in the Advanced Disease setting, while the first item is Hazard Ratio (HR) in the Adjuvant setting.  
However, as shown in Section \ref{sec.HRnotLogical}, HR being not Logic-respecting can lead to wrong targeting of patients.  
Therefor we recommend against using HR.  
Instead, we recommend using Time Ratio (TR) to assess efficacy in OS, since \cite{Ding&Lin&Hsu(2016)} has shown TR to be Logic-respecting.  

\section{Transitioning out of Hazard Ratio (HR)}

For the FDA to transition from one system to another, it is essential to establish a bridging mechanism providing continuity between legacy and the new system, ensuring that therapeutic products previously deemed effective under the prior framework remain approvable and do not fail solely due to methodological changes, and safeguarding against the inadvertent approvability of products that were previously determined to be ineffective. 
From my experience helping Bioequivalence transition from legacy \textit{testing-for-difference} to \textit{two 1-sided t-tests} (\cite{Hsu&Hwang&Liu&Ruberg(1994)}, \cite{Berger&Hsu(1996)}), setting the equivalence margin of (0.8, 1.25) by re-analyzing data from the legacy system harmonized the assessments and ensured transitional comparability.  
With it, the FDA maintained regulatory consistency and preserved stakeholder confidence throughout the system migration.

\subsection{Weibull distribution modeling smooths transitioning}

Cancer cells often use multiple pathways simultaneously to resist treatment.  
So the Weibull distribution is frequently suitable for modeling OS, since it arises as the limiting distribution of the \emph{minimum} of a sample from a continuous distribution.  
Indeed, in \cite{ReidCox(1994)}, Sir David Cox (of the Cox PH model) said ``I would normally want to tackle problems parametrically, so I would take the underlying hazard to be a Weibull or something.''  

The Weibull distribution is both an Accelerated Failure Time (AFT) model and a Cox Proportional Hazards (PH) model, with coefficients directly connecting Time Ratio (TR) to Hazard Ratio (HR).  
The FDA has previous oncology approval data which it can re-analyze to get a sense of the magnitude of TR that corresponds to prior approvals based on HR, for a smooth transition.  
In transitioning from Last Observation Carried Forward (LOCF) to Mixed-Effect Model Repeated Measure (MMRM) analysis in a different therapeutic area, \cite{SiddiquiHung(2009)} indeed re-analyzed 25 neurological and psychiatric NDAs.  

However, for the results from such retrospective analyses to not be misled by how the analyses are stratified differently, the mistake in computer packages need to be fixed first.  

\subsection{Subgroup Mixable Estimation fixes issue in computer packages}

The first principle of Subgroup Mixable Estimation (SME) is Nature mixes probabilities, by prevalence.  
For example, if 40\% of females respond while 70\% of males respond, with a 50:50 mix of female:male patients, the overall population has 55\% responders.  

Hazard Ratios (HRs) cannot be mixed directly, because they are not probabilities.  
To correctly mix HRs 
so that stratified HR for the overall population will agree with the conditional HRs (after the Prognostic effect is taken into account), they have to be converted to probability equivalents.  

A meaningful probability in survival analysis is the probability that, among all possible random pairing of patients in which one is given $Rx$ the other givens $C$, the one given $Rx$ lives longer than the one given $C$, what \cite{LiuWangTianHsu(2022)} call the Living Longer Probability (LLP).  
Test statistic of the \cite{Mann-Whitney1947} test for comparing outcome under $Rx$ and $C$ is an estimate of LLP, and one can think of the null hypothesis it tests as $LLP = \frac{1}{2}$.   

Motivation for Enrich Lehmann to study in \cite{Lehmann(1953)} the family of distributions that has since been named after him was, if the family of survival distributions is in a Lehmann family (which would now be termed as having Proportional Hazard, PH for short), then \textit{power} of the Mann-Whitney test does not depend on the parametric form of the distribution, but only on the exponent that the reference distribution is raised to, which Sir David Cox later defined as HR in \cite{Cox1972}.  

\cite{LiuWangTianHsu(2022)} showed that, for a family of survival curves with PH, HR is the Odds of Living Shorter.  
They further developed a correspondence between HR and LLP, which is $LLP = \frac{1}{1+HR}$.  
Living Longer Probabilities (LLPs) are probabilities, so they can be mixed by prevalence, as shown in equation (2) of \cite{LiuWangTianHsu(2022)}.  

Accordingly, in research with Jiaqian Liu and Ying Ding at University of Pittsburgh, to correctly mix HRs, we transform HRs to LLPs, mix LLPs to obtain LLP for the overall population, then transform back to obtain the HR for the overall population as demonstrated in Table \ref{tab:Watermelon}.  
Our software that fixes the oversight in computer packages carefully follows a second principle of SME, which is the Prognostic effect should always be taken into account in assessing efficacy (regardless of whether the efficacy measure is logic-respecting or not).  

\section{Concluding recommendation}

Logicality requires, and Subgroup Mixable Estimation (SME) delivers, an efficacy assessment for the overall population within the range of minimum and maximum efficacy in the subgroups, no matter how outcome is measured, whichever logic-respecting efficacy measure is chosen, the same efficacy assessment regardless of how subgroups are stratified.  

\section{Acknowledgment}
Logicality, the noun for being Logic-respecting, was suggested by Xingya Wang, who also suggested and made the plot for Figure \ref{fig:logicRespecting}.  

\bibliographystyle{Chicago}
\bibliography{Freferences}
\end{document}